\documentstyle[12pt]{article}

\catcode`@=11
\long\def\@caption#1[#2]#3{\par\addcontentsline{\csname
  ext@#1\endcsname}{#1}{\protect\numberline{\csname
  the#1\endcsname}{\ignorespaces #2}}\begingroup
    \small
    \@parboxrestore
    \@makecaption{\csname fnum@#1\endcsname}{\ignorespaces #3}\par
  \endgroup}
\catcode`@=12

\input{epsf}
\ifx\epsffile\undefined
\newlength{\epsfysize}
\def\epsffile#1#2#3#4]#5{}
\fi
\input{rotate}

\renewcommand{\theequation}{\thesection.\arabic{equation}}
\newcommand{\be}{\begin{equation}}   
\newcommand{\ee}{\end{equation}}
\newcommand{\bear}{\begin{eqnarray}}
\newcommand{\eear}{\end{eqnarray}}
\newcommand{\ba}{\begin{array}}      
\newcommand{\ea}{\end{array}}
\newcommand{\lae}{\begin{array}{c}\,\sim\vspace{-21pt}\\< \end{array}}
\newcommand{\gae}{\begin{array}{c}\,\sim\vspace{-21pt}\\> \end{array}}
\newcommand{\eL}{{\cal L}}
\newcommand{\Q}{{\cal Q}}
\newcommand{\R}{{\cal R}}
\newcommand{\m}{{\tilde m}}
\begin{document}

\pagestyle{empty}
\begin{titlepage}
\def\thepage {}        

\title{\bf A Chiral Supersymmetric Standard Model\\ [1cm]}

\author{ {\small \bf Hsin-Chia Cheng, 
Bogdan A.~Dobrescu and Konstantin T.~Matchev} \\ 
\\
{\small {\it Fermi National Accelerator Laboratory}}\\
{\small {\it P.O. Box 500, Batavia, IL 60510, USA \thanks{e-mail
  addresses: hcheng@fnal.gov, bdob@fnal.gov, matchev@fnal.gov} }}\\ }

\date{ }

\maketitle

   \vspace*{-9.4cm}
\noindent
\makebox[10.7cm][l]{ hep-ph/9807246 } FERMILAB-PUB-98/205-T \\ [1mm]
\makebox[10.7cm][l]{\today} \\ 

 \vspace*{10.5cm}

\baselineskip=18pt

\begin{abstract}

   {\normalsize
We propose a supersymmetric extension of the Standard Model with an
extra $U(1)$ gauge symmetry, so that all supersymmetric
mass terms, including the $\mu$-term,
are forbidden by the gauge symmetries.
Supersymmetry is broken dynamically which results in $U(1)$ breaking and 
generation of realistic $\mu$ term and soft breaking masses.
The additional fields required to cancel the $U(1)$ anomalies are 
identified with the messengers of supersymmetry breaking. The gaugino 
masses arise as in the usual gauge mediated scenario, while squarks
and sleptons receive their masses from both the $U(1)$ $D$-term and the 
two-loop gauge mediation contributions. The scale of supersymmetry 
breaking in this model can be below $10^6$ GeV, yielding collider 
signatures with decays to goldstinos inside the detector.}

\end{abstract}

\vfill
\end{titlepage}

\baselineskip=18pt
\pagestyle{plain}
\setcounter{page}{1}

\section{Introduction}

The main theoretical shortcoming of the Standard Model (SM) is the lack of 
chirality of the Higgs sector: the gauge symmetry does not prevent a 
large mass for the Higgs doublet, which results in the hierarchy problem.
Supersymmetry (SUSY) confers chirality to scalars, and therefore 
offers the hope of explaining the hierarchy between the electroweak scale 
and the Planck scale. However, in the Minimal Supersymmetric Standard
Model (MSSM), the Higgs sector is still not 
chiral: the gauge symmetry allows a $\mu$ term of order the Planck
scale. Furthermore, most realistic models with dynamical 
SUSY breaking require gauge singlets and additional vector-like fields.

Given the current lack of understanding of quantum gravity, the only rigorous
way of avoiding dangerously large masses for singlet or vector-like
representations, is to introduce
gauge symmetries which forbid any linear or quadratic term in the 
superpotential. Since at least one of the Higgs doublets must carry the
new charges, the simplest choice for the additional gauge group is a $U(1)$.

The known examples of $U(1)$ gauge symmetries that prevent a large 
$\mu$ term \cite{mu} have been constructed in the framework of supergravity 
mediated SUSY breaking. However, in order to give 
large enough gaugino masses, the dynamical SUSY breaking sector 
has to include a 
gauge singlet which couples to the gauge superfields~\cite{gaugino}.
Therefore, such models are not chiral.

Thus, we are led to consider a gauge mediated SUSY breaking scenario
\cite{DNS}.
The $U(1)$ gauge group has to be anomaly free, so that the model 
must include extra fields which are vector-like under the SM gauge
group and charged under the $U(1)$.
These fields may naturally play the role of the messenger sector.
In this letter we construct a complete, renormalizable and calculable
model of gauge mediation, including an explicit 
dynamical SUSY breaking (DSB) sector. 
We use the new $U(1)$ gauge symmetry both to forbid the $\mu$ term
and to communicate SUSY breaking to the SM superpartners.
As a result, the scale of $U(1)$ breaking and the $\mu$ 
parameter are related to the scale of SUSY breaking.
The model has no pure gauge singlets, nor vector-like representations,
i.e., it is a purely chiral supersymmetric Standard Model.
 
\section{Model Building}

Our starting point is, within the MSSM, to forbid the $H_u H_d$ term in the 
superpotential 
by charging the two Higgs superfields under a new $U(1)_\mu$ gauge group.
In order to have a Higgsino mass, the usual $\mu$ term of the Higgs sector
must come from an $S H_u H_d$ term in the 
superpotential, where $S$ 
(which is a $SU(3)_C\times SU(2)_W \times U(1)_Y$ singlet, but
charged under $U(1)_\mu$) acquires a vacuum expectation value (vev) 
at or below the $U(1)_\mu$ breaking scale.

The Yukawa couplings of the Higgs doublets require the quarks and leptons 
to be charged under $U(1)_\mu$ too. Therefore, some
extra fields charged under the $U(1)_\mu$ and SM gauge groups
will in general be required to cancel the arising anomalies. 
These extra fields can play the role of 
the messenger sector used in gauge mediation \cite{DNS}. This sector 
includes some chiral superfields, $q, \bar{q}, l, \bar{l}$, transforming
non-trivially under the SM gauge group, and superpotential
terms $X q \bar{q}$, $X l \bar{l}$, where $X$ 
is a SM singlet whose scalar and $F$ components acquire vevs
$\langle X\rangle$ and $\langle F_X\rangle$.
Note that the relevant $q \bar{q}$, $l \bar{l}$, $X$ and $X^2$ operators in 
the superpotential are forbidden by the $U(1)_\mu$.

The fact that $X$ and the messengers carry $U(1)_\mu$ charges
implies that this symmetry is broken at a scale of order
$\langle X\rangle$ or higher,
so that the $U(1)_\mu$ $D$-term is expected to be significant. 
As a result, the $D$-term
contributions to squark and slepton masses may dominate or be comparable to
the usual two loop contributions mediated by the SM gauge interactions.
Since the $D$-term contributions to the squared masses are proportional to the 
scalar charges, we need to give same sign $U(1)_\mu$ charges to all the quarks 
and leptons. 

Apparently, it is non-trivial to cancel the $U(1)_\mu$ anomalies in this case. 
The MSSM has $[SU(3)_C]^2 \times U(1)_\mu$, $[SU(2)_W]^2 \times U(1)_\mu$, 
$U(1)_Y^2 \times U(1)_\mu$ and $U(1)_Y \times U(1)_\mu^2$ anomalies, which 
should be cancelled by the anomalies of the messengers. 
A simple solution is to observe that $E_6 \supset SU(5)_{\rm SM}
\times U(1)_\mu$, where $SU(5)_{\rm SM}$ includes the SM gauge group.  
The fundamental representation of $E_6$ decomposes as ${\bf 27} = 
({\bf 10 + \bar{5} + 1} , +1) + ({\bf 5 + \bar{5}}, -2) + ({\bf 1}, +4)$, 
and is anomaly free.
Motivated by this, although we do not require 
$E_6$ gauge unification, we introduce (in addition to the MSSM fields,
which include the right handed neutrinos)
three $q, \bar{q} \in ({\bf 3 + \bar{3}})$ of $SU(3)_C$, 
and two $l, \bar{l} \in ({\bf 2 + 2})$ of $SU(2)_W$.
Then, we can assign $U(1)_\mu$ charge +1 to all the quarks and leptons;
and charge $-2$ to $H_{u,d}$ as well as to all the messengers.
This ensures that the $U(1)_\mu$ anomalies mentioned above cancel.
In addition, the $S$ and $X$ fields
can be identified with two of the three $({\bf 1}, +4)$  representations of 
$SU(5)_{\rm SM} \times U(1)_\mu$ required for $U(1)_\mu^3$ 
anomaly cancellation.

In order to allow $X$ to acquire an $F$ term, we include a new 
SM singlet, $N$, with $U(1)_\mu$ charge $-2$, and a $X N^2$ term
in the superpotential. The complete field content of the MSSM + messenger sector is given in Table~\ref{tab:mssm}.
One can check that indeed the gauge symmetries
do not allow any supersymmetric mass term. 
The renormalizable superpotential that we consider includes, in addition to the 
usual couplings of $H_{u,d}$ to quarks and leptons, only the following terms
\be
\label{superpotential}
W = f_q X q_i{\overline q}_i + f_l X l_j{\overline l}_j
+ \frac{\lambda}{2} XN^2 - \frac{\epsilon}{2} SN^2 + \kappa SH_uH_d . 
\ee
%
\begin{table}[ht]
\centering
\renewcommand{\arraystretch}{1.4}
\begin{tabular}{|c||c|c|c||c|}\hline
Fields $i=1,2,3$
  & $SU(3)_C$
    & $SU(2)_W$
      & $U(1)_Y$
        & $U(1)_\mu$ \\
\hline \hline
$Q_{i}$
  & $3$
    & 2
      & $+{1\over6}$
        & $+1$
\\ \hline
$U_{i}, D_{i}$
  & ${\overline 3}$
    & 1
      & $-{2\over3}, +{1\over3}$
        & $+1$
\\ \hline
$L_{i}$
  & 1
    & 2
      & $-{1\over2}$
        & $+1$
\\ \hline
$E_{i}, \nu_{i}$
  & 1
    & 1
      & $+1, 0$
        & $+1$
\\ \hline\hline
$H_u, {\overline l}_1, {\overline l}_2 $
  & 1
    & 2
      & $+{1\over2}$
        & $-2$ \\ \hline 
$H_d, l_1,l_2$
  & 1
    & 2
      & $-{1\over2}$
        & $-2$
\\ \hline 
$q_{i}$
  & $3$
    & 1
      & $+{1\over3}$
        & $-2$
\\ \hline 
${\overline q}_i$
  & ${\overline 3}$
    & 1
      & $-{1\over3}$
        & $-2$
                  \\ \hline 
$X,S$
  & 1
    & 1
      & $0$
        & $+4$
                     \\ \hline
$N$
  & 1
    & 1
      & $0$
        & $-2$
                     \\ \hline
\end{tabular}
\parbox{5.5in}{
\caption{Particle content and charge assignments for the MSSM+messenger sector.
\label{tab:mssm}}}
\end{table}

Apart from the  MSSM + messenger sector described so far, there should be
a sector that breaks SUSY dynamically. The DSB
sector also contains fields 
which transform under $U(1)_\mu$, so that $U(1)_\mu$ can play the role
of the messenger group which communicates SUSY
breaking to the visible sector.
The properties of this DSB sector are constrained
by the low energy structure we introduced so far.
The first condition is that the $U(1)_\mu$ and 
$U(1)_\mu^3$ anomalies of the MSSM + messenger sector, 
given by $(-4)+(-2)$ and $(-4)^3 + (-2)^3$,
have to be cancelled by the $U(1)_\mu$ and $U(1)_\mu^3$ anomalies of 
the DSB sector. Another condition is that 
the supertrace of the $U(1)_\mu$ charged fields in the DSB sector 
is positive, so that 
the $X$ and $S$ scalars receive negative squared masses and acquire vevs.
In addition, we have to make sure that after $U(1)_\mu$ is broken
by the $N$, $S$, $X$ vevs, the $U(1)_\mu$ $D$-term contributions
to the squared masses of the squarks and sleptons are positive.

An example of a DSB sector satisfying our requirements is
the $SU(4)\times SU(3)$
model described in Ref.~\cite{PST}. The field content is shown
in Table~\ref{tab:dsb}. One can check that all anomalies cancel in the
combination of the MSSM + messenger sector and the DSB sector. 
After including the superpotential which
lifts the flat directions, it breaks SUSY dynamically.
The SUSY breaking minimum is discussed in the Appendix. For the
analysis of the visible sector we only need to know that the DSB
sector generates a negative squared mass for each scalar charged
under $U(1)_\mu$, proportional to its $U(1)_\mu$ charge squared,
and a contribution to the $U(1)_\mu$ $D$-term, $-\xi^2$, (which
is unimportant as long as $\xi$ is much smaller than the $U(1)_\mu$
breaking scale).
\begin{table}[ht]
\centering
\renewcommand{\arraystretch}{1.4}
\begin{tabular}{|c||c|c||c|}\hline
Fields
	  & $SU(4)$
            & $SU(3)$ 
	& $U(1)_\mu$\\
\hline \hline
${\cal Q}$
          & 4
            & 3 & $-{1\over 2}$ \\ \hline
${\cal L}_1, \, {\cal L}_2, \, {\cal L}_3$
          & ${\overline 4}$
            & 1 
        & $+{5\over 2}, \, +{1\over 2}, \, -{3\over 2}$ \\ \hline
${\cal R}_1, {\cal R}_2, \R_3,\R_4$
          & 1
            & ${\overline 3}$
        & $-2, \, 0, \, +2, +2$ \\ \hline
\end{tabular}
\parbox{5.5in}{
\caption{Particle content and charge assignments in the DSB sector.
\label{tab:dsb}}}
\end{table}
Then, the relevant part of the potential for the MSSM $+$ messenger sector
is given by
\bear
V & = &
\frac{g_\mu^2}{2} \left(-\xi^2+4|X|^2+4|S|^2-2|N|^2-2|H_u|^2 - 2|H_d|^2
+...\right)^2 \qquad\qquad
\nonumber \\
&-& \m^2\left(16 |X|^2+16|S|^2+4|N|^2 + 4|H_u|^2 + 4 |H_d|^2 +...\right)
+  \frac{\lambda^2}{4} |N|^4 
\nonumber \\
&+&   
\left|\frac{\epsilon}{2} N^2-\kappa H_u H_d \right|^2
+ |N|^2 \left|\lambda X-\epsilon S\right|^2 
+ \kappa^2|S|^2
\left( |H_u|^2+|H_u|^2 \right) + ... ~,
\eear
where the ellipsis stand for terms involving squarks, sleptons, and messenger 
scalars. 
The values of $\xi$ and $\m$ are given in the Appendix as 
functions of the parameters in the DSB sector.
In Section 3 it is shown that the constraints on the gaugino masses and on the
$B$ and $\mu$ parameters from the Higgs sector lead to
$\lambda^{3/2} < \epsilon \ll \lambda \ll 1$. 
We assume that the hierarchy of these couplings, as well as 
the hierarchy of fermion masses come from some unknown physics at high 
scales\footnote{A supersymmetric model of flavor utilizing
an extra gauge $U(1)$ for gauge mediation
was just recently proposed in \cite{Elazaar}.}.

Minimizing the potential is straightforward, and for
$\kappa > \sqrt{\lambda^2 +\epsilon^2}$  we find a desired
minimum at $\langle H_u \rangle = \langle H_d \rangle = 0$ and 
\be
\langle N^2 \rangle ={24\m^2\over\lambda^2+\epsilon^2},  \quad
\langle X \rangle = {\epsilon\over\lambda}\langle S \rangle, \quad
\langle S^2 \rangle =
{\lambda^2 \over \lambda^2+\epsilon^2}
\left(
{\xi^2 \over4}
+{ \m^2\over g_\mu^2}
+{12\m^2\over\lambda^2+\epsilon^2}
\right) ~.
\ee
This is only a local minimum, but we expect that its lifetime 
is sufficiently long. We comment more on this in Section 4.
The corresponding SUSY-breaking $F$ and $D$-terms are given by 
\be
\langle F_N \rangle =0, \quad
\langle F_X \rangle = \frac{\lambda}{2} \langle N^2 \rangle \simeq
\sqrt{6} \m \langle N \rangle, \quad
\langle F_S \rangle = -\frac{\epsilon}{2}  \langle N^2 \rangle, \quad
g_\mu^2 \langle D \rangle = 4 \m^2.
\ee
The $\langle X \rangle$ and $\langle F_X \rangle$ vevs provide
the SUSY preserving and breaking masses for the messenger fields,
$q,\bar{q}, l, \bar{l}$, while $\langle S \rangle$ and $\langle F_S \rangle$
provide the $\mu$ and $B$ term for the Higgs sector.
Gaugino masses come from the usual one-loop gauge mediation contribution.
The scalar squared masses receive, in addition to the usual two-loop
SM gauge mediation contributions, a $U(1)_\mu$ $D$-term contribution
and a negative contribution from the $U(1)_\mu$ mediation.
For superpartner masses near the weak scale, we require
$\m \lae 1\; \mbox{TeV}$, $\sqrt{F_X} \gae 30$ TeV,
hence the $U(1)_\mu$ breaking scale $\langle N \rangle \gae
10^3$ TeV and $\lambda \lae 10^{-3}$.
The resulting sparticle spectrum is discussed in the next
Section.

\section{Sparticle spectrum}

The communication of SUSY breaking from the DSB sector to
the visible sector proceeds in two steps. First, at the scale
\be
M_\mu\equiv g_\mu \langle N \rangle \simeq 
2\sqrt{3} {g_\mu\over \lambda}\tilde m \quad (\gg \m)
\label{Mmu}
\ee
of $U(1)_\mu$ breaking, each scalar with $U(1)_\mu$ charge $Q^{f}_\mu$ 
which does not acquire a vev receives a soft mass contribution
\be
m_{\tilde f}^2(M_\mu) 
= Q^{f}_\mu (4-Q^{f}_\mu) \tilde m^2.
\ee
In particular, all squarks and sleptons get a positive
squared mass $+3\tilde m^2$, while the two Higgs doublets get
a negative soft squared mass of $-12\tilde m^2$.
The $\mu$ and $B$ terms are also generated at this scale:
\be
\mu (M_\mu)= \kappa \langle S \rangle 
\simeq 2\sqrt{3} {\kappa \over \lambda}\tilde m \quad (\gae \tilde m),
\label{mu}
\ee
\be
B (M_\mu)= {\langle F_S\rangle \over \langle S\rangle} \simeq
-2\sqrt{3}{ \epsilon \over \lambda}\tilde m 
\quad(|B|\ll \tilde m).
\label{B}
\ee
One may check that the condition $\kappa>\sqrt{\lambda^2 +\epsilon^2}$ 
ensures that the tree-level squared masses for the Higgs doublets are positive 
and thus electroweak symmetry is unbroken at this stage.

The messenger fermions and scalars also become massive at the scale
$M_\mu$. The masses of the $d_i,{\overline d}_i$ and
$l_i,{\overline l}_i$ fermion messengers are
\be
M_{q,l}\equiv f_{q,l} \langle X\rangle \simeq 
2\sqrt{3}{\epsilon f_{q,l}\over\lambda^2}\tilde m \quad (\gg \m).
\label{Mql}
\ee
The messenger scalars, on the other hand, have the
following mass matrices:
\be
{\cal M}^2_{\tilde q,\tilde l} \approx M_{q,l}^2 
\left[ {\bf 1} + \frac{\lambda^3}{\epsilon^2 f_{\tilde q,\tilde l}} 
\left( \ba{cc} - \lambda/f_{\tilde q,\tilde l} & 1 
\\ 1 & - \lambda/f_{\tilde q,\tilde l} \ea \right) \right] ~.
\ee
The eigenvalues are positive if $\epsilon > {\cal O}(\lambda^{3/2})$,
assuming $f_{q,l} \sim {\cal O}(1)$.

Finally, the gauge singlets $X$, $S$ and $N$
also get masses. Their fermionic components mix among themselves, and
with the $U(1)_\mu$ gaugino.
As a result, we find two Dirac fermions, with masses of order
$24(g_\mu/\lambda)\tilde m$ and $4\tilde m$, respectively.
The scalar components of the singlets also mix, and the resulting mass spectrum
is as follows. There is a massless Nambu-Goldstone boson, which is
eaten by the $U(1)_\mu$ gauge boson; and there is a
scalar of mass $24(g_\mu/\lambda)\tilde m$, which becomes a member
of the heavy gauge supermultiplet. The rest of the scalars
are light, with masses $2\sqrt{6}\tilde m, 2\sqrt{6}\tilde m,
2\sqrt{3}\tilde m$ and $2\sqrt{2}\tilde m$, correspondingly.

Below the messenger scale, $M\simeq M_q \simeq M_l$, the messengers
$d_i,{\overline d}_i$ and $l_i,{\overline l}_i$ are integrated out
inducing gaugino masses
\be
M_n(M) = c_n{\alpha_n\over4\pi}\Lambda g\left(\Lambda/ M\right),
\ee
where $n = 1,2,3$ corresponds to $U(1)_Y$, $SU(2)_W$ and $SU(3)_C$,
$c_1 = 4, c_2 = 2, c_3 = 3$ (we do not use the $SU(5)$
 normalization for $\alpha_1$), and
\be
\Lambda \equiv {\langle F_X \rangle \over \langle X\rangle}
\simeq 2\sqrt{3}{\lambda\over\epsilon}\tilde m ~.
\label{Lambda}
\ee
The messengers also contribute to the scalar masses,
\be
m^2_{\tilde f}(M)={2\Lambda^2\over(4\pi)^2}
\left( 3C_3^{f}\alpha_3^2 +2C_2^{f}\alpha_2^2
+\frac{20}{3}C_1^{f}\alpha_1^2 \right) f\left(\Lambda/M\right) ~,
\label{msq}
\ee
where the coefficients $C_i^{f}$ are zero for gauge singlet sfermions
$\tilde f$, and $4/3$, $3/4$
and $Y^2$ ($Y=Q-I_3$ denotes the usual hypercharge)
 for fundamental representations of $SU(3)_C$,
$SU(2)_W$ and $U(1)_Y$, correspondingly.  The threshold functions
$f(x)$ and $g(x)$ can be found in Ref.~\cite{fg}. 
In order to get large enough gaugino masses, we need
$\Lambda \sim 30\; \mbox{TeV} \gg \m$, hence $\epsilon \ll \lambda$.
In eq.~(\ref{msq})
we neglect small contributions of order 
$[\alpha/(4\pi)^2]^2 \tilde m^2 \ln(M_\mu/M)$, 
arising due to the nonvanishing messenger supertrace \cite{EPST}.

The effect of the renormalization group equations (RGEs)
on the sparticle spectrum between the $M_\mu$ and $M$ scales is peculiar:
due to their larger Yukawa couplings,
the third generation sfermions are driven {\em heavier} than those
of the first two generations. The reason is that
in our model the soft squared masses for the Higgs doublets are negative
due to the $U(1)_\mu$ $D$-term, and as a result, the scalar mass
combinations
\be
M_{Q_3}^2 + M_{U_3}^2 + M_{H_u}^2 \simeq
M_{Q_3}^2 + M_{D_3}^2 + M_{H_d}^2 \simeq
M_{L_3}^2 + M_{E_3}^2 + M_{H_d}^2 \simeq
 -6\tilde m^2 ~,
\ee
which appear in the RGEs, are also negative. Of course, below the messenger scale 
the usual gauge mediated contributions are included, and their Yukawa RGE 
effect on the third generation soft masses is just the opposite. The net effect
depends on the values of the model parameters. Note that the charge
assignments are such that ${\rm Tr}\,Q_Y\cdot Q_\mu=0$, so that a
hypercharge $D$-term is not induced through RGE renormalization.

The supersymmetric mass spectrum at the electroweak scale is determined
as a function of the following parameters:
$\{\tilde m, \kappa/\lambda, \epsilon/\lambda, f_{q,l}/\lambda, 
g_\mu/\lambda \}$.
Using eqns.~(\ref{Mmu}), (\ref{Lambda}), (\ref{mu})-(\ref{Mql}), 
and the requirement for radiative electroweak
symmetry breaking,
they can be exchanged for $\{\Lambda, M, M_\mu, \tan\beta, {\rm sign}(\mu)\}$.
Note the presence of the extra parameter $M_\mu$ as compared to the minimal 
gauge mediated models. The only constraint on the parameter space is
$M_\mu\gg M>\Lambda$. We would also expect to be in the large
$\tan\beta$ region, since our model predicts low values of $B$.
\begin{figure}[t]
\epsfysize=3.5in
\epsffile[-40 220 320 600]{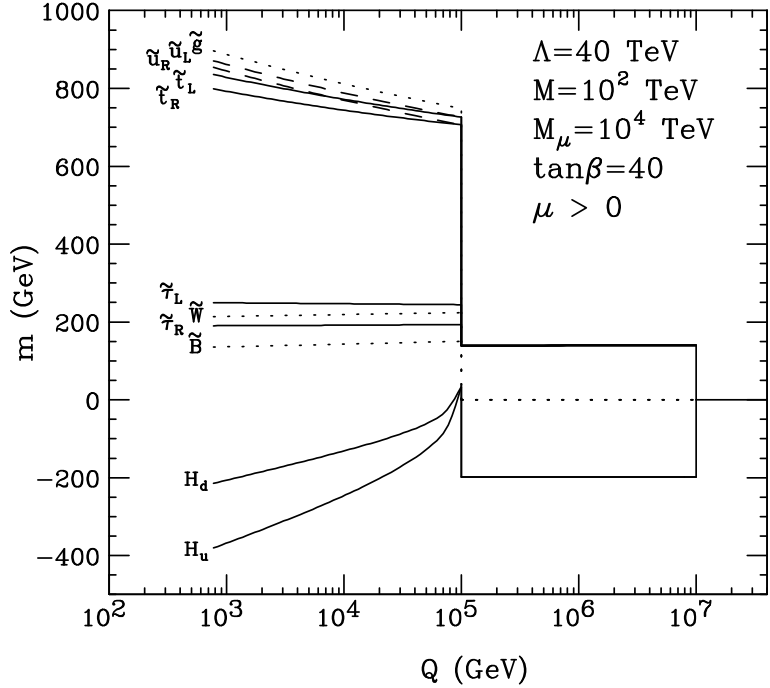}
\begin{center}
\parbox{5.5in}{
\caption[RGE evolution.]
{\small RGE evolution of the soft mass parameters: gaugino masses
(dotted lines), first two generation squark masses (dashed) and
stop, stau and Higgs masses (solid). Since the soft Higgs masses can be
negative, we plot ${\rm sign}(m^2)\sqrt{|m^2|}$.
\label{fig1}}}
\end{center}
\end{figure}
In Fig.~\ref{fig1} we show the RGE evolution of representantive soft masses
for $\Lambda=40$ TeV, $M=10^2$ TeV, $M_\mu=10^4$ TeV, $\tan\beta=40$ and
$\mu>0$. We find that these values correspond to $\m=80$ GeV,
$\kappa/\lambda=1.5$, 
$\epsilon/\lambda=7\times 10^{-3}$, $f_{q,l}/\lambda=5.1\times 10^4$ and 
$g_\mu/\lambda=3.6\times 10^4$. 

Qualitatively, the superpartner spectrum looks similar to 
the pure gauge mediation scenario, if $\m$ is smaller than
the SM gauge mediated soft masses. Then, as
$\m$ gets larger than the SM gauge mediation contributions, the
squarks and sleptons become heavier in comparison to the gauginos,
and increasingly degenerate, since they carry the same $U(1)_\mu$ charge.

\section{Discussion}\label{sec:discussion}

Several features of the model presented here warrant further comments.

{\bf 1.}  It is well known that the ``$\mu$ problem'' is more difficult
to solve in gauge mediation models. Besides the general problem of a potentially
large $\mu$-term induced by Planck scale physics, 
a totally separate sector is often required solely for generating the
$\mu$-term \cite{DNS,musol}. 
If one tries to generate the $\mu$ term directly
from the messenger sector by a small coupling or a
loop contribution, the resulting ratio of $B\mu$ and $\mu$ is too 
big\footnote{However, there are ways to cure this problem \cite{muprob}.}, 
$B\mu/\mu \sim F_X/X \sim 10^4-10^5$ GeV.
In our model, the supersymmetric Higgs mass term is forbidden
by the $U(1)_\mu$ gauge symmetry, and the $\mu$ term
is generated only after $U(1)_\mu$ is broken. The
$S$ field which gives rise the $\mu$-term is an integral
part of this model and is required for anomaly cancellation.
The $F_S/S$ ratio is small, allowing acceptable values 
for $B\mu /\mu$. However, without understanding
the small Yukawa couplings ($\lae 10^{-3}$) needed in this
model, we can not claim that the scale of the $\mu$ term
is completely natural.

{\bf 2.}  The intrinsic SUSY breaking scale in this model can be
as low as $10^5-10^6$ GeV, lower than that in the original models
of gauge mediation~\cite{DNS}. To see this, we assume that the
$U(1)_\mu$ breaking scale ($\langle N \rangle$, determined by
$\m$ and $\lambda$) is about the same as the scale of heavy fields
in the DSB sector, $\langle \R \rangle$, so that
the formulae (\ref{D-term}), (\ref{2loopmass}) in the appendix
can still apply. Then we have 
\be
F_X 
\approx \sqrt{6} \m \langle N \rangle 
\sim c \sqrt{480}
\left( \frac{g_\mu^2}{16\pi^2} \right) m_b \langle \R \rangle
\sim c' \sqrt{480} \left( \frac{g_\mu^2}{16\pi^2} \right)
E_{\rm vac}^2,
\ee
where $c,\; c'$ are ${\cal O}(1)$ constants, $E^4_{\rm vac}$ is the 
vacuum energy density, and $m_b$ is defined in (\ref{mb}).
We can see that the intrinsic SUSY breaking
scale $E_{\rm vac}$ has to be only about less than an order of magnitude
higher than the messenger scale. For $\sqrt{F_X}\sim 10^4-10^5$ GeV,
$E_{\rm vac}$ can be lower than $10^6$ GeV.
In that case, the next to lightest supersymmetric
particle (NLSP) can decay inside the detector, yielding
interesting collider signals~\cite{NLSP}. Note that most gauge
mediation models have the SUSY breaking scale so high that NLSP
will escape the detector, giving similar
signals as in the traditional supergravity mediation scenario.
All previously known models with the SUSY breaking scale below $10^6$
GeV so far involve some assumptions about noncalculable strong 
dynamics \cite{LSSB}, therefore it is not certain that they are viable.

{\bf 3.}  The vev of the $N$ field can give Majorana masses to the
right-handed neutrinos on the order of the $U(1)_\mu$ breaking scale
via superpotential interactions $N \nu_i \nu_j$. However, without
knowing the Yukawa couplings of the neutrinos, we are not able to
predict the neutrino masses and mixing patterns.

{\bf 4.} We did not include all possible terms consistent with the
gauge symmetry in the superpotential (\ref{superpotential}).
Since we allow small Yukawa couplings, most of the missing couplings
could be sufficiently small, so that they do not have significant
effects on our model. Some of them may change the low energy
parameters. For example, a small coupling between $X$ and 
$H_u, H_d$ can give extra contribution to the $B\mu$ parameter, and
hence affect $\tan \beta$.
The only dangerous terms are those matter-messenger couplings
which induce proton decays, so we may need an extra symmetry to
forbid them (the flavor changing constraints
are not very severe \cite{HZ}). 
A more attractive solution in this framework would be to
have different messenger fields or different charge assignments
so that the messenger-matter couplings which allow rapid proton
decays can not exist.

{\bf 5.} The $U(1)_\mu$ gauge symmetry forbids $R$ parity violating operators.
Moreover, the $U(1)_\mu$ is broken only by fields with even charges, such that 
a $Z_2$ symmetry, identified as the $R$ parity, is automatically conserved.

{\bf 6.}  As in the original models of gauge mediation \cite{DNS},
the minimum we considered is a local minimum \cite{ACHM,DDR}.
There exist lower minima and even runaway directions with
$\langle X \rangle=0, \; \langle \bar{q} q \rangle 
\mbox{ and/or } \langle \bar{l} l \rangle \neq 0$.
Ref.~\cite{DDR} estimates the vacuum tunneling rate. It is
found that in order for the lifetime of the local minimum
to be longer than the age of the universe, some Yukawa couplings
have to be small ($\lambda < 0.1$), which is easily satisfied
in our case.

To our knowledge, the model we have presented here is the first example
of a purely chiral supersymmetric Standard Model.
This is a viable model and yields interesting superpartner
spectrum and phenomenology, which may be tested in future experiments.

\vspace{.25in}

{\it Acknowledgements:} We would like to thank Jon Bagger,
Joe Lykken and Sandip Trivedi for useful discussions. 
Fermilab is operated by URA under DOE contract DE-AC02-76CH03000.

\appendix
\setcounter{equation}{0}
\renewcommand{\theequation}{A.\arabic{equation}}


\section{The DSB Sector}

An interesting set of DSB models is the $SU(N)\times SU(N-1)$
models of Poppitz, Shadmi and Trivedi~\cite{PST}. One can require
that the superpotential preserves an $SU(N-2)$ global symmetry.
If we add a fundamental representation of the $SU(N-2)$
(singlet under $SU(N)\times SU(N-1)$), it is anomaly free~\cite{AMM}. 
In addition, there is a non-anomalous $U(1)$, so that 
we can gauge a $U(1)$ subgroup of $SU(N-2)\times U(1)$ and
use it as the messenger to transmit SUSY breaking to the visible sector.

The model which we use is the $SU(4)\times SU(3)$ model. The
field content of the DSB sector  and their charges under the
``messenger'' $U(1)_\mu$ are shown in Table~\ref{tab:dsb}. 
The $U(1)_\mu^3$ and $U(1)_\mu$ anomalies (which would be cancelled
by adding two fields with charges $-4$, $-2$) are cancelled by the
combination of the MSSM fields and the messenger sector.

The superpotential of the DSB sector is given by
\be
W_{DSB}\ =\ 
  \lambda_1\eL_1\Q\R_1
+ \lambda_2\eL_2\Q\R_2
+ \lambda_3\eL_3\Q\R_3
+ {\alpha\over3!}\R_1\R_2\R_4 .
\ee
For this model to be calculable, we assume that $\alpha \ll
\lambda_1, \; \lambda_2, \; \lambda_3 \; \sim 1$, so that
the vacuum lies in the weakly coupled regime. The detailed
analysis of this family of models can be found in Ref.~\cite{AMM}.
Here we just sketch the result. The $\R_i$ fields develop
large vevs and give large masses
to the $\eL_i$ and $\Q$ fields. After integrating out the heavy
fields, the low energy nonlinear SUSY sigma model is described 
by the baryons $b_i$, where
\be
b_i = \frac{1}{3!} \epsilon_{ijkl} \R_j \R_k \R_l ,
\ee
with a superpotential
\be
W_{\rm eff}= (\Lambda_D^9 b_4)^{\frac{1}{4}} + \alpha b_3 .
\ee
The first term comes from the gaugino condensation of the
$SU(4)$ gauge group (with scale $\Lambda_4$).
We have absorbed the $\lambda_i$ couplings into $\Lambda_D$, 
$\Lambda_D^9 \sim \lambda_1 \lambda_2 \lambda_3 \Lambda_4^9$.
Combining it with the K\"{a}hler potential,
\be
K_{\rm eff}=3 (b_i^\dagger b_i)^{1/3} ,
\ee
we find that the minimum occurs at
\be
b_3 = -0.075 (\alpha^{-\frac{4}{9}} \Lambda_D)^3 ,\quad
b_4 = 0.102  (\alpha^{-\frac{4}{9}} \Lambda_D)^3 .
\ee
The energy density at the minimum and the masses of the
scalar components of $b_1, \; b_2$ are
\bear
E^4_{\rm vac} &=& 0.220 \alpha^{\frac{2}{9}} \Lambda_D^4 ,\\
m_{b_{1,2}}^2 \equiv m_b^2 &=& 0.445 \alpha^{\frac{10}{9}} \Lambda_D^2 .
\label{mb}
\eear
One can easily see that $E^2_{\rm vac}\sim m_b \langle \R \rangle$.
Because the light fields $b_1$ and $b_2$ have $U(1)_\mu$ charges
$+4$ and $+2$ respectively, they will generate the Fayet-Illiopoulos
$D$ term for the $U(1)_\mu$ gauge group as discussed in Ref.~\cite{DNS},
\be
-\xi^2 = - \sum_j \frac{g_\mu^2}{16\pi^2} Q_\mu^{b_j} m_{b_j}^2
\ln \frac{M_V^2}{p^2} 
= -6 \left( \frac{g_\mu^2}{16\pi^2}\right)  m_{b}^2
\ln \frac{M_V^2}{p^2} ,
\label{D-term}
\ee
where $M_V$ represents the mass scale of the heavy fields
in the DSB sector, and $p^2$ is the larger scale between the
$U(1)_\mu$ breaking scale, $M_\mu^2$, and $m_b^2$.
They also generate a negative contribution to the mass squared
of each scalar field charged under $U(1)_\mu$ at two-loop, proportional
to the field's charge squared,
\be
\frac{m_i^2}{q_i^2} \equiv -\m^2 = -\sum_{j} 4 \left(
\frac{g_\mu^2}{16\pi^2}\right)^2 \left(Q_\mu^{b_j}\right)^2 
m_{b_j}^2 \ln \frac{M_V^2}{p^2}
= -80 \left(\frac{g_\mu^2}{16\pi^2}\right)^2 m_{b}^2 
\ln \frac{M_V^2}{p^2} .
\label{2loopmass}
\ee
Note that the formulae (\ref{D-term}), (\ref{2loopmass}) only
apply when $p^2 < M_V^2$. If the $U(1)_\mu$ breaking scale ($p^2=M_\mu^2$)
is higher than $M_V^2$, the results will be suppressed by a
factor $M_V^2/M_\mu^2$.


\vfill

\end{document}